\documentclass[dvips]{article}
\usepackage{color}
\usepackage{stfloats}
\fnbelowfloat
\usepackage{url} 
\usepackage[switch,pagewise]{lineno}
\usepackage{graphicx,amssymb,amsmath,times,epsfig}
\usepackage{units}
\usepackage{nicefrac}
\usepackage{booktabs}
\usepackage{icrc2011}
\def\us{\char`\_} 

\def\la{\mathrel{\mathchoice {\vcenter{\offinterlineskip\halign{\hfil
$\displaystyle##$\hfil\cr<\cr\sim\cr}}}
{\vcenter{\offinterlineskip\halign{\hfil$\textstyle##$\hfil\cr<\cr\sim\cr}}}
{\vcenter{\offinterlineskip\halign{\hfil$\scriptstyle##$\hfil\cr<\cr\sim\cr}}}
{\vcenter{\offinterlineskip\halign{\hfil$\scriptscriptstyle##$\hfil\cr<\cr
\sim\cr}}}}}

\def\ga{\mathrel{\mathchoice {\vcenter{\offinterlineskip\halign{\hfil
$\displaystyle##$\hfil\cr>\cr\sim\cr}}}
{\vcenter{\offinterlineskip\halign{\hfil$\textstyle##$\hfil\cr>\cr\sim\cr}}}
{\vcenter{\offinterlineskip\halign{\hfil$\scriptstyle##$\hfil\cr>\cr\sim\cr}}}
{\vcenter{\offinterlineskip\halign{\hfil$\scriptscriptstyle##$\hfil\cr>\cr
\sim\cr}}}}}

\title{Highlights from the Pierre Auger Observatory}

\newcommand{\etal}{\MakeLowercase{\textit{et al. }}} 
\shorttitle{Kampert \etal Auger Highlights}

\authors{Karl-Heinz Kampert$^{1}$ for the Pierre Auger Collaboration$^{2}$}
\afiliations{$^1$Department of Physics, University of Wuppertal, Germany\\ $^2$Observatorio Pierre Auger, Av. San Mart\'in Norte 304, 5613 Malarg\"ue, Argentina \\(Full author list: http://www.auger.org/archive/authors\texttt{\us}2011\texttt{\us}05.html)}
\email{auger\texttt{\us}spokespersons@fnal.gov}

\abstract{
This paper summarizes some highlights from the Pierre Auger Observatory that were presented at the ICRC 2011 in Beijing. 
The cumulative exposure has grown by more than 60\,\% since the previous ICRC to above 21\,000 km$^2$\,sr\,yr. Besides giving important updates on the energy spectrum, mass composition, arrival directions, and photon- and neutrino upper limits, we present first measurements of the energy spectrum down to $3 \cdot 10^{17}$\,eV, first distributions of the shower maximum, $X_{\rm max}$, together with new surface detector related observables sensitive to $X_{\rm max}$, and we present first measurements of the p-air cross section at $\sim 10^{18}$\,eV. Serendipity observations such as of atmospheric phenomena showing time evolutions of elves extend the breadth of the astrophysics research program.}


\begin{document}
\maketitle

\section{Introduction and Status of the Observatory}

The Pierre Auger Observatory started collecting data in 2004. Since the completion of the base-line Observatory in 2008 its aperture has grown by about 7000 km$^2$\,sr, each year. At this meeting we present data based on an exposure of more than 21\,000 km$^2$\,sr\,yr. The Auger Observatory uses hybrid measurements of air showers recorded by an array of 1660 water Cherenkov surface stations covering an area of 3000\,km$^2$, together with 27 air fluorescence telescopes that observe the development of air showers in the atmosphere above the array during dark nights. 

An infill array with half the grid size has been completed and we present first data at this meeting extending the energy spectrum down to $3\cdot 10^{17}$\,eV, thereby covering the ankle of the primary energy spectrum with full detection efficiency. Moreover, the three high-elevation telescopes (HEAT) started operation and - together with the infill array in the FOV of the telescopes - will allow us to extend the hybrid measurements further down to $10^{17}$\,eV with unprecedented precision. This will enable the study of the transition from galactic to extra-galactic cosmic rays. Construction of the buried muon detectors (AMIGA) in the in-fill area is in progress. Measurements of the muons are important for studying the composition of cosmic rays from surface detector data and their information is also of vital importance for studying particle interactions down to the energy of the LHC. In addition, an extensive R\&D program for radio detection of UHE air showers is under way and construction of the Auger Engineering Radio Array (AERA) has started. Once completed, it will comprise 160 radio antennas distributed over an area of 20\,km$^2$. First triple hybrid events composed of particle densities at ground, longitudinal shower profiles from fluorescence telescopes, and radio signals from the first antennas have already been observed. Last but not least, an intense R\&D program for microwave detection of air showers has begun with the first GHz-antennas presently being installed. These extensions and new technologies may enhance the performance and capabilities of the Auger Observatory in Argentina and, in parallel, will explore their potential for a future much larger ground based observatory.

There are 38 papers presented on behalf of the Pierre Auger Collaboration at this meeting, and they are all accessible in five e-print compilations with arXiv numbers 1107.4804, 1107.4805, 1107.4806, 1107.4807, and 1107.4809.

\section{The energy spectrum}
\label{sec:espec}

An accurate measurement of the cosmic ray flux above $10^{17}$ eV is crucial for discriminating between different models describing the transition between galactic and extragalactic cosmic rays, the suppression induced by the cosmic ray propagation, and the features of the injection spectrum at the sources. Two complementary techniques are employed at the Pierre Auger Observatory: a surface detector array (SD) and a fluorescence detector (FD). The energy spectrum at energies greater than $3 \cdot 10^{18}$\,eV has been derived using data from the SD-array. The analysis of air showers measured with the FD which also triggered at least one station of the surface detector array (i.e.\ hybrid events) enables measurements to be extended to lower energies. Despite the limited number of events, due to the fluorescence detector on-time, the lower energy threshold and the good energy resolution of hybrid events allow us to measure the flux of cosmic rays with the standard array down to $10^{18}$\,eV, into the energy region where the transition between galactic and extragalactic cosmic rays is expected.

The energy calibration of the SD-array is based on so-called golden hybrid events, i.e.\ events that can be independently reconstructed from the SD and FD data. Applying high quality cuts, 839 events could be used for the SD calibration \cite[R.\ Pesce]{ICRC-Comp-I}. The overall FD energy resolution is 7.6\,\% and it is almost constant with energy. The total systematic uncertainty on the FD energy scale is about 22\,\%. It includes contributions from the absolute fluorescence yield (14\,\%), calibration of the fluorescence telescopes (9.5\,\%), the invisible energy correction (4\,\%), systematics in the reconstruction method used to calculate the shower longitudinal profile (10\,\%), and atmospheric effects (6\,\% - 8\,\%). The atmospheric uncertainties include those related to the measurements of aerosol optical depth (5\,\% - 7.5\,\%), phase function (1\,\%) and wavelength dependence (0.5\,\%), the atmosphere variability (1\,\%) and the residual uncertainties on the estimation of pressure, temperature and humidity dependence of the fluorescence yield (1.5\,\%).
 
\begin{figure}[!t]
\centering
\includegraphics[width=.48\textwidth]{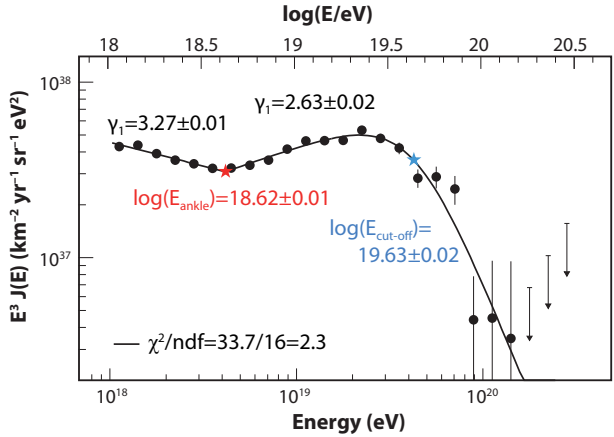}
\vspace*{-2mm}\caption{\small{Combined energy spectrum fitted with two power laws in the ankle region and a smoothly changing function at higher energies. Only statistical uncertainties are shown. The systematic uncertainty in the energy scale is 22\,\%.}}
\label{fig:espec}
\end{figure}

The energy spectrum derived from hybrid data has been combined with the one obtained from surface detector data using a maximum likelihood method and is shown in Fig.\,\ref{fig:espec} together with a broken power law and a smooth cut-off at higher energies \cite[F.\ Salamida]{ICRC-Comp-I}. Both, the ankle and suppression of the flux at higher energies are clearly visible. The spectrum can be compared to astrophysical models and can be described by both a proton and heavy-dominated composition at the highest energies. Thus, measurements of the composition are needed to discriminate between various astrophysical models.

Data of the 750\,m infill array reach full efficiency for all primaries at $E>3\cdot 10^{17}$\,eV and, using data with an exposure of 26 km$^2$\,sr\,yr, extend the spectrum of Fig.\,\ref{fig:espec} smoothly down to this threshold \cite[I.\ Mari\c{s}]{ICRC-Comp-I}. Analysis of the composition in this energy range is on-going. HEAT data, combined with the infill array, extend the energy range further down to $10^{17}$\,eV \cite[H.J.\ Mathes]{ICRC-Comp-V}.

\section{The cosmic ray mass composition}
\label{sec:compos}

\begin{figure}[!t]
\centering
\includegraphics[width=.4\textwidth]{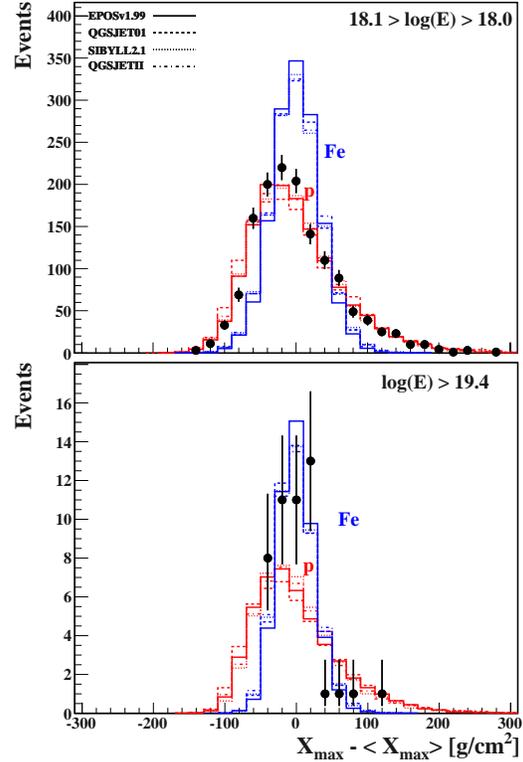}
\vspace*{-2mm}\caption{\small{Centered distribution, $X_{\rm max}-\langle X_{\rm max} \rangle$, for the lowest and highest energy bins. Subtraction of the mean allows the comparison of the shapes of these distributions with the superimposed MC simulations.}}
\label{fig:xmax-dist}
\end{figure}

As discussed above, measuring mass composition of cosmic rays along with the flux is a key to separating the different scenarios of origin and propagation of cosmic rays. The composition must be inferred from measurements of various shower observables, most importantly the atmospheric depth at which the shower attains its maximum size, $X_{\rm max}$. For a given shower, the position of $X_{\rm max}$ will depend on the depth of the first interaction of the primary in the atmosphere and the depth that it takes the cascade to develop. Thereby, it will depend not only on the primary mass, but also on the cross section of the primary particle with air and on features of hadronic interactions at high energies. This important caveat should be kept in mind when discussing the mass composition of cosmic rays, i.e.\ interpretation of shower observables in terms of primary mass are subject to deficiencies of hadronic interaction models employed in air shower simulations. Besides the position of $X_{\rm max}$, its fluctuations on shower-by-shower basis, RMS$(X_{\rm max})$, show strong sensitivity to the primary mass. 

\begin{figure}[!t]
\centerline{
\includegraphics[width=.4\textwidth]{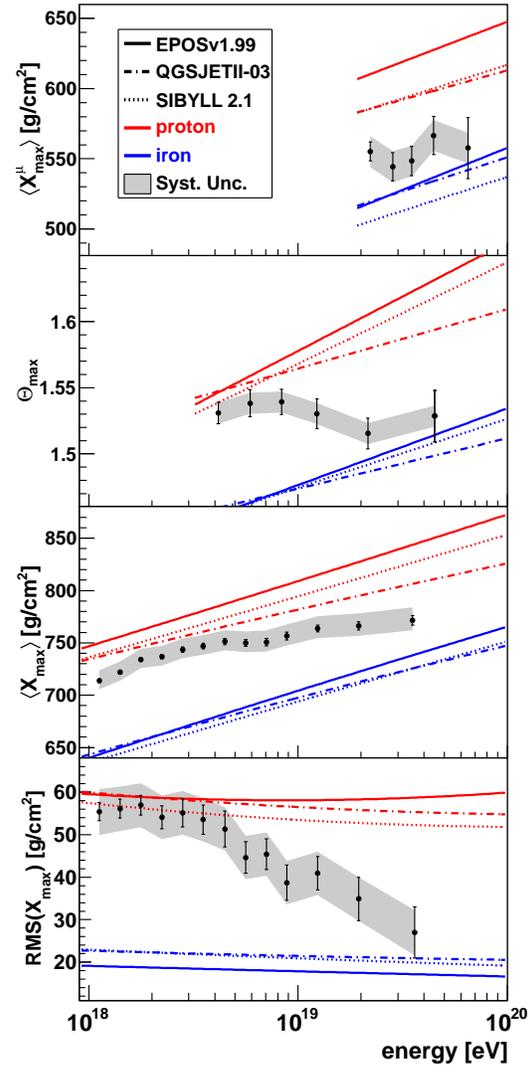}}
\vspace*{-2mm}\caption{{\small Results on shower evolution sensitive observables compared with models prediction. The error bars correspond to the statistical uncertainty. The systematic uncertainty is represented by the shaded bands.}}
\label{fig:mass-compilation}
\end{figure}

For the analysis, again hybrid data are used and the shower profiles are required to be good fits to a Gaisser-Hillas function, as deviations could indicate the presence of residual clouds. Both $\langle X_{\rm max} \rangle$ and its RMS show a characteristic change at $E \ga 5\cdot 10^{18}$\,eV indicating an increasingly heavier composition when compared to EAS simulations \cite[P.\ Facal]{ICRC-Comp-II} (c.f.\ Fig.\,\ref{fig:mass-compilation}). It is well known that MC predictions are more uncertain for the $\langle X_{\rm max} \rangle$ than for the fluctuations. This is mainly due to the additional dependence of $\langle X_{\rm max} \rangle$ on the multiplicity in hadronic interactions. In Fig.\ \ref{fig:xmax-dist} we therefore compare the shape of the distributions, $X_{\rm max}-\langle X_{\rm max} \rangle$, to MC predictions for different compositions and hadronic interaction models. As can be seen, in this representation the various models predict a nearly universal shape. At low energy, the shape of the distribution is compatible with a very light or mixed composition, whereas at high energies, the narrow shape would favor a significant fraction of nuclei (CNO or heavier).

Fluorescence telescopes are the only observational tool currently enabling direct measurements of the shower maximum $X_{\rm max}$. Unfortunately, those data suffer from statistics because of the duty cycle being only $\sim 15$\,\%. However, surface detectors, operated 24 hours a day, also provide observables which are related to the longitudinal shower profile. These observables are subject to independent systematic uncertainties (both experimentally and theoretically). The higher statistics allow us to extend these measurements to higher energies than possible with the FD.

For each SD event, the water-Cherenkov detectors record their signals as a function of time. Since muons travel in almost straight lines whereas the electromagnetic particles suffer more multiple scattering on their way to ground, the first part of the signal is dominated by
the muon component. Due to the absorption of the electromagnetic (EM) component, the number of these particles at the ground depends,
for a given energy, on the distance to the shower maximum and
therefore on the primary mass. In consequence, the time profile of
particles reaching ground is sensitive to the cascade development: the
higher the production height, the narrower the time pulse.
The time distribution of the SD signal is characterised by means of the risetime, $t_{1/2}$, which depends on the distance to the shower
maximum, the zenith angle $\theta$ and the distance to the core
$r$. In a first step, the zenith angle at which the risetime asymmetries between the inner and outer SD stations of a shower become maximal is is analyzed. This angle $\Theta_{\rm max}$ is then, in a second step, related to the shower maximum using a subset of hybrid events \cite[D.\ Garcia-Pinto]{ICRC-Comp-II}. Using this correlation it is possible to measure the shower evolution with surface detector data, in a similar way to that in which the SD energy calibration is performed for a subset of events with the FD data. The result is shown in Fig.\,\ref{fig:mass-compilation}.

Not only the risetime of signals in the SD tanks, but also the arrival time of particles with respect to the shower front plane contains information about the position of the shower maximum. A method for reconstructing the so-called Muon Production Depth (MPD), i.e.\ the depth at which a given muon is produced, measured parallel to the shower axis, using the FADC traces of detectors far from the core, has been presented in~\cite{cazon_2004}. From the MPDs an observable can be defined, $X_{\rm max}^\mu$, as the depth along the shower axis where the number of produced muons in a shower reaches a maximum. The method is currently restricted to inclined showers where muons dominate the signal at ground level. Once the MPDs are obtained for each event, the value of $X_{\rm max}^\mu$ is found by fitting a Gaisser-Hillas function to the depth profile. The results of $\langle X_{\rm max}^\mu\rangle$ presented in Fig.\,\ref{fig:mass-compilation} are restricted to zenith angles between $55^\circ$ and $65^\circ$ and use timing information only for detectors far from the core ($r >  1800$\,m). Because of this distance restriction, the effective energy range for which the method can presently be applied is limited to $E > 2\cdot 10^{19}$\,eV. The measured values of $\langle X_{\rm max}^\mu \rangle$ are presented in the upper panel of Fig.~\ref{fig:mass-compilation}.  It is important to point out, that the predictions of $X^\mu_{\rm max}$ from different hadronic models would not be affected if a discrepancy between a model and data~\cite[J.\ Allen]{ICRC-Comp-II} is limited to the total number of muons. However, differences in the muon energy and spatial distribution would modify the predictions.

With this caveat concerning hadronic interaction models, one might infer the primary composition from the data on the longitudinal air shower development presented in Fig.\,\ref{fig:mass-compilation}. The evolution of $\langle X_{\rm max}\rangle$, $\Theta_{\rm max}$ and $\langle X_{\rm max}^\mu\rangle$  with energy is similar, despite the fact that the three analyses come from completely independent techniques that have different sources of systematic uncertainties.  Concerning the RMS of $X_{\rm max}$, a variety of compositions can give rise to large values of the RMS, because the width of the $X_{\rm max}$ is influenced by both, the shower-to-shower fluctuations of individual components and their relative displacement in terms of $\langle X_{\rm max}\rangle$ \cite{kampert-unger-12}. However, within experimental uncertainties, the behaviour of $\langle X_{\rm max}\rangle$,  $\Theta_{\rm max}$ and $\langle X_{\rm max}^\mu\rangle$  as shown in  Fig.~\ref{fig:mass-compilation} is compatible with the energy evolution of RMS($X_{\rm max}$). In particular, at the highest energies all four analyses show consistently that our data better resemble the simulations of heavier primaries than pure protons.

\section{p-air cross section and tests of hadronic interaction models}
\label{sec:int-models}

One of the biggest challenges for a better understanding of the nature of ultra-high energy cosmic rays is to improve the modeling of hadronic interactions in air showers. None of the current models is able to consistently describe cosmic ray data, which most importantly prevents a precise determination of the primary cosmic ray mass composition.

\begin{figure}[bt]
\centering
\includegraphics[width=.45\textwidth]{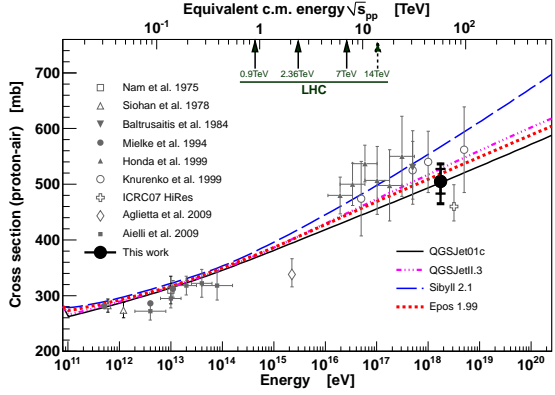}
\vspace*{-.3cm}
\caption{\label{fig:p-air}{\small Proton-air cross section compared to other measurements and model predictions (for references see~\cite[R.\ Ulrich]{ICRC-Comp-II}). The inner error bars are statistical only, while the outer include all systematic uncertainties for a helium fraction of 25\,\% and 10\,mb photon systematics.}}
\end{figure}

Studies to exploit the sensitivity of cosmic ray data to
the characteristics of hadronic interactions at energies beyond
state-of-the-art accelerator technology began over $50$\,years ago. The property of interactions most directly linked to the development of extensive air showers is the cross-section for the production of hadronic particles~(e.g.~\cite{Ulrich:2010rg}). To reconstruct the proton-air cross-section based on hybrid data, we analyse the shape of the distribution of the largest values of the depth of shower maxima, $X_{\rm max}$.  This \emph{tail} of the $X_{\rm max}$-distribution, 
that contains the $20\,\%$ of deepest showers, exhibits the expected exponential shape $dN/dX_{\rm max} \propto \exp{-(X_{\rm max}/\Lambda_{20})}$. It is directly related to the p-air cross section via $\sigma_{\rm p-air} = \langle m_{\rm air} \rangle / \Lambda_{20}$. In practice, to properly account for shower fluctuations and detector effects, the exponential tail is compared to Monte Carlo predictions. Any disagreement between data and predictions is then attributed to a modified value of the proton-air cross-section \cite[R.\ Ulrich]{ICRC-Comp-II}. In this analysis, the energy interval is restricted to $10^{18}$ to $10^{18.5}$\,eV which corresponds to a center-of-mass energy in the nucleon-nucleon system of $\sqrt{s}=57$\,TeV. This interval has been chosen because of high statistics in the data and because of the composition being compatible with a dominance of protons (see. Sec.\,\ref{sec:compos}). A possible contamination of He primaries could mimic a larger cross section (e.g.\ by 20\,mb for 20\,\% He contamination) while a photon contamination could reduce the cross section by at most 10\,mb. Combining the results one finds
\begin{equation*}
  \sigma_{\rm p-air} = \left(505\;\pm{22_{\rm stat}}\; \;
  (_{-14}^{+19})_{\rm syst}  \right)\;\rm{mb}
\end{equation*}
at a center-of-mass energy of $57\pm6$\,TeV. This result is shown in comparison to other data and models in Fig.\,\ref{fig:p-air}.

The result favors a moderately slow rise of the cross section towards
higher energies, well in line with recent results from LHC (e.g.\ \cite{Aad:2011ct}). A conversion of the derived $\sigma_{\rm p-air}$ measurement into the more fundamental cross-section of proton-proton collisions using the Glauber framework~\cite{Glauber:1955qq,Glauber:1970jm} will be published elsewhere.\\

\begin{figure}[t]
\centering
\includegraphics [width=.45\textwidth]{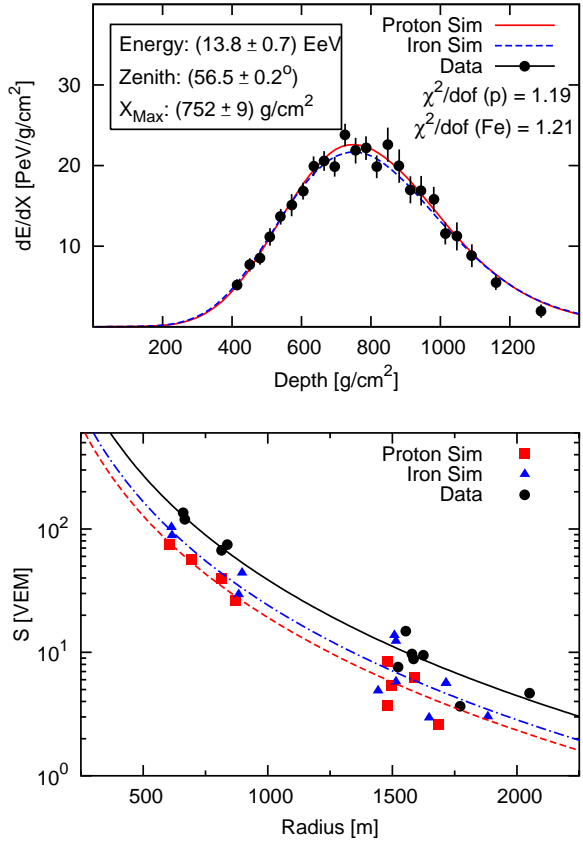}
\vspace*{-2mm}\caption{{\small Top: A longitudinal profile measured for a hybrid event and matching simulations of two showers with proton and iron primaries. Bottom: A lateral distribution function determined for the same hybrid event as in the top panel and that of the two simulated events.}}
\label{fig:TD-LPLDFComp}
\end{figure}

The importance of hadronic interaction models to measurements of the cosmic ray mass composition has been addressed in Sect.\,\ref{sec:compos}. In particular, muons in extensive air showers are subject to large theoretical uncertainties due to our limited knowledge of multi-particle production in hadronic interactions. However, hybrid data can be used to constrain the models and to uncover deficiencies in describing features of EAS data. When measuring the longitudinal profile (LP) in a golden hybrid event, we construct a library of simulated air-shower events with the same shower geometry where the LP of each simulated event matches a measured one. The measured LP constrains the natural shower-to-shower fluctuations of the distribution of particles at ground. This allows the ground signals of simulated events to be compared to the ground signals of measured events on an event-by-event basis. An example of such an analysis is shown in Fig.\,\ref{fig:TD-LPLDFComp}~\cite[J.\ Allen]{ICRC-Comp-II}. Here, the LP of a measured event is compared to p and Fe simulations, each providing a good fit to the data. The bottom panel shows the corresponding signals in the SD.  The ratio of the measured signal at 1000\,m from the shower core, $S(1000)$, to that predicted in simulations of showers with proton primaries, $\frac{S(1000)_{\text{Data}}}{S(1000)_{\text{Sim}}}$, is 1.5 for vertical showers and grows to around 2 for inclined events. The ground signal of more-inclined events is muon-dominated. Therefore, the increase of the discrepancy with zenith angle suggests that there is a deficit of muons in the simulated showers compared to the data. The discrepancy exists for simulations of showers with iron primaries as well, which means that the ground signal cannot be explained only through composition.

\begin{figure}[!t]
\centering
\includegraphics[width=0.45\textwidth]{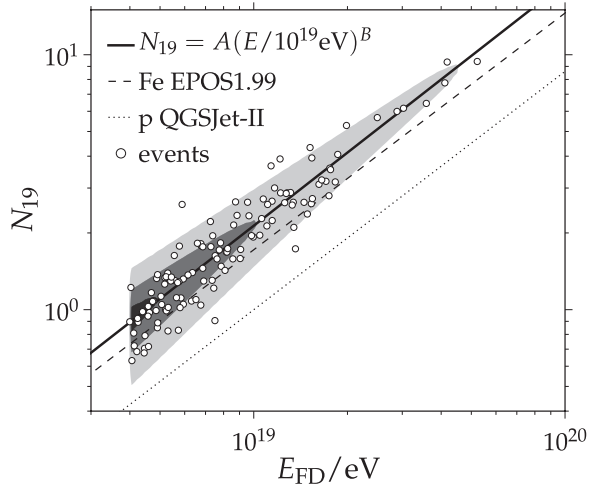}
\vspace*{-2mm}\caption{{\small Fit of a calibration curve $N_{19} = A(E/10~{\rm EeV})^B$. The constants $A$ and $B$ are obtained using the maximum-likelihood method. The contours indicate the constant levels of the p.d.f.\ integrated over zenith angle, corresponding to 10, 50 and 90\% of the maximum value~\cite[H.\ Dembinski]{ICRC-Comp-I}. Calibration curves for protons QGSJETII (dot line) and iron EPOS1.99 (dashed line) are shown for comparison.}}
\label{fig:Calib}
\end{figure}

This finding is corroborated by direct estimations of the muon number from the signal traces in the SD as well as by making use of universality of the muonic to electromagnetic signal $S_\mu(1000)/S_{\rm em}(1000)$ for fixed vertical depth of the shower \cite[J.\ Allen]{ICRC-Comp-II}. Moreover, this purely-observational estimation of the muonic signal in data, is compatible with results obtained from inclined showers~\cite[G.\ Rodriguez]{ICRC-Comp-II}. This is best illustrated by Fig.\,\ref{fig:Calib}. Here, $N_{19}$ is defined as the ratio of the total number of muons, $N_{\mu}$, in the shower with respect to  the total number of muons at $E=10$\,EeV given by a 2-dim reference distribution, $N_{19}=N_{\mu}(E,\theta) / N^{{\rm map}}_{\mu}(E=10~{\rm EeV},\theta)$ which accounts for the geomagnetic spatial deviation of muons at ground.

Thus, all of the analyses show a significant deficit in the number of muons predicted by simulations with proton primaries compared to data. This discrepancy cannot be explained by the composition alone, although a heavy composition could reduce the relative excess by up to 40\,\%. The increased sophistication of the methods gives further weight to the conclusions that, at the current fluorescence energy scale, the number of muons in data is nearly twice that predicted by simulations of proton-induced showers. The possible zenith angle dependence of $N_{\mu}^{\text{rel}}$ suggests that, in addition to the number, there may also be a discrepancy in the attenuation and lateral distribution of muons between the simulations and data.

\section{Update of photon and neutrino upper limits}
\label{sec:photons-neutrinos}

The search, and possibly study, of high energy photons and neutrinos is of interest for at least three reasons: (i) Top-Down models of UHECR origin \cite{bhattacharjee99} including topological defect or super-heavy dark matter models predict a significant fraction of photons and neutrinos at the highest energies, (ii) they would provide a smoking-gun signature of the GZK-effect because of the decay of charged and neutral pions created in photo-pion production, and (iii) they would open a new window to the most extreme Universe by possibly seeing point sources in the sky. A search for their signatures has thus been part of the Auger research program from the beginning. 

The search for EeV photons presented at this meeting is based on hybrid events. Due to the FD duty cycle the event statistics is reduced compared to the SD-only detection mode. However, the hybrid detection technique provides a precise geometry and energy determination with the additional benefit of allowing to reduce the energy threshold for detection to about $10^{18}$\,eV.  To improve the photon-hadron discrimination power over measurements of  $X_{\rm max}$ only, the differences in the lateral distribution functions for photons and hadrons measured by the  SD have been considered by analyzing the observable, $S_{b}$, defined in~\cite{Roos-11}. To reject misreconstructed profiles, only time periods with the sky not obscured by clouds, and with a reliable measurement of the vertical optical depth of aerosols, are selected. On the SD side we require at least 4 active stations within 2~km of the hybrid reconstructed axis. This prevents an underestimation of $S_{b}$ (which would mimic the behavior of a photon event) due to missing or temporarily inefficient detectors. For the classification of photon candidates, a Fisher analysis trained with a sample of a total of $\sim$30000 photon and proton CORSIKA~\cite{corsika} showers generated according to a power law spectrum between $10^{17}$ and $10^{20}$~eV is performed~\cite[M.\ Settimo]{ICRC-Comp-III}. The Fisher response distributions for photon and proton primaries are well separated for all energies above $10^{18}$\,eV. Photon-like events in the data are then selected by applying an ``a priori'' cut to the upper 50\,\% of the photon like events. This reduces the photon detection efficiency to 50\,\% but provides a conservative result in the upper limit calculation by reducing the dependence on the hadronic interaction models and on the mass composition assumption. With this choice, the expected hadron contamination is about 1\,\%  in the lowest energy interval (between 10$^{18}$ and 10$^{18.5}$~eV) and it becomes smaller for increasing energies~\cite[M.\ Settimo]{ICRC-Comp-III}.
 
\begin{figure}[!t]
\centering
\includegraphics[width=0.45\textwidth]{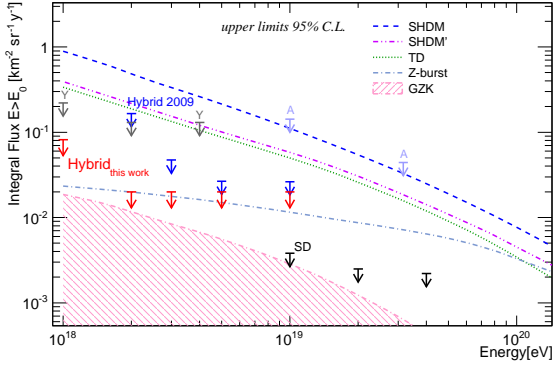}
\vspace*{-2mm}\caption{{\small Upper limits on the photon flux above 1,~2,~3,~5 and 10 EeV derived in this work (red arrows) compared to previous limits from Auger, from AGASA (A), and Yakutsk (Y). The shaded region and the lines give the predictions for the GZK photon flux and for top-down models (TD, Z-Burst, SHDM and SHDM'). (See \cite[M.\ Settimo]{ICRC-Comp-III} for references.)}}
\label{fig:photon-limit}
\end{figure}

Applying the method to data, 6,~0,~0,~0 and~0 photon candidates are found for energies above 1, 2, 3, 5 and 10~EeV. We checked with simulations that the observed number of photon candidates is consistent with the expectation for nuclear primaries, under the assumption of a mixed composition. The corresponding 95\,\% CL upper limits on the photon flux $\Phi_{\gamma}^{95CL}$ integrated above an energy threshold $E_{0}$ are shown in Fig.\,\ref{fig:photon-limit}. To be conservative, a minimum value of the exposure above $E_{0}$ is used and a possible nuclear background is not subtracted for the calculation of $N^{95CL}_{\gamma}$. The flux limits shown in Fig.\,\ref{fig:photon-limit} or likewise the derived limits on the photon fraction of 0.4\%, 0.5\%, 1.0\%, 2.6\% and 8.9\% above 1,~2,~3,~5 and 10~EeV, significantly improve previous results at the lower energies and rule out exotic models of UHECR origin, except for the $Z$-burst model of Ref.\,\cite{gelmini}. While the focus of the current analysis was the low EeV range, future work will be performed to improve the photon-hadron separation also at higher energies using further information provided by the SD.\\

The surface detector is well suited also to search for ultra-high energy neutrinos in the sub-EeV energy range and above. Neutrinos of all flavours can interact in the atmosphere and induce inclined showers close to the ground (down-going). The sensitivity of the SD to tau neutrinos is further enhanced through the ``Earth-skimming" mechanism (up-going). Both types of neutrino interactions can be identified through the broad time structure of the signals induced in the SD stations. 

The analysis starts with the inclined shower selection (down-going: $\theta > 75^{\circ}$ and Earth-skimming $\theta<96^{\circ}$). These showers usually have elongated patterns on the ground along the azimuthal arrival direction. A length $L$ and a width $W$ are assigned to the pattern and a cut on their ratio $L/W$ is applied. We also calculate the apparent speed $V$ of an event using the times of signals at ground and the distances between stations projected onto $L$. Finally, for down-going events, we reconstruct the zenith angle $\theta_{\rm rec}$. After this pre-selection, the FADC signal-traces of the SD stations are analyzed to search for so-called ``young showers'' with a broad time structure. To optimize the discrimination power, again a Fisher discriminant method is used and trained to a subset of data. The identification efficiency for the set of selection cuts applied to the data depends on the neutrino energy $E_\nu$, the slant depth $D$ from ground to the neutrino interaction point, the shower geometry, the neutrino flavour ($\nu_e$, $\nu_\mu$, or $\nu_\tau$), and is different for CC- and NC-type interactions, see~\cite{Auger-neutrino-11}.

\begin{figure}[!t]
\centering
\includegraphics[width=.45\textwidth]{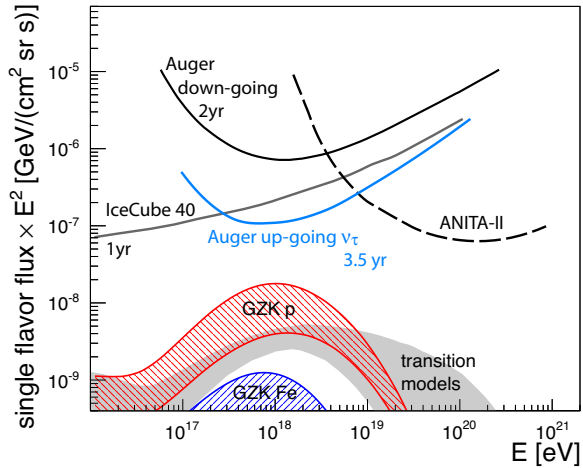}
\vspace*{-2mm}\caption{{\small Differential upper limits (90$\%$ C.L. per half decade of energy) from the Pierre Auger Observatory for a diffuse flux of down-going $\nu$ (2 yr of full Auger) and Earth-skimming $\nu_\tau$ (3.5 yr of full Auger \cite[Y.\ Guardincerri]{ICRC-Comp-III}). Limits from other experiments and expected fluxes are also shown (see \cite{kampert-unger-12} for references).}}\label{fig:neutrino}
\end{figure}

Using the independent sets of identification criteria that were designed to search for down- and up-going neutrinos in the data collected from 1 January 2004 to 31 May 2010, no candidate was found ~\cite[Y.\ Guardincerri]{ICRC-Comp-III}.  Assuming a differential flux $f(E_{\nu}) = k E_{\nu}^{-2}$, we place a 90\% CL upper limit on the single flavour neutrino flux of  $k<3.2 \times 10^{-8}~{\rm GeV~cm^{-2}~s^{-1}~sr^{-1}}$ in the energy interval $1.6\times10^{17}~{\rm eV} - 2.0\times10^{19}~{\rm eV}$, based on Earth-skimming neutrinos and $k<1.7 \times 10^{-7}~{\rm GeV~cm^{-2}~s^{-1}~sr^{-1}}$ in the energy interval $1\times10^{17}~{\rm eV} - 1\times10^{20}~{\rm eV}$, based on down-going neutrinos (see Fig.\,\ref{fig:neutrino}). The optimistic fluxes for p-primaries shown in this figure are accessible for the proposed lifetime of the Pierre Auger Observatory. The transition models and sources with a dominance of heavy primaries would be challenging to reach for any of the currently operating experiments.

With no candidate events found in the search period, we can also place a limit on the UHE neutrino flux from a source at declination $\delta$. Since the sensitivity to UHE$\nu$s is limited to large zenith angles, the rate of events from a point source in the sky depends strongly on its declination. In both Earth-skimming and down-going analyses the sensitivity yields a broad ``plateau'' spanning $\Delta\delta\sim 110^\circ$ in declination with the highest sensitivity reached at $\delta \simeq \pm 55^\circ$. The present flux limits do not yet allow us to constrain models of UHE$\nu$ production in the jets and  the core of CenA~\cite{theoreticalPrediction_CenA}.

\section{Anisotropies}
\label{sec:anisotropies}

One of the keys to understanding the nature of UHECRs is their distribution over the sky. This distribution depends on the location of the UHECR sources, as well on the UHECR mass composition and large-scale magnetic fields, both Galactic and extragalactic. Despite significant efforts, none of these issues is well understood at present. Observation of the suppression of the CR flux at the highest energies (c.f.\ Sect.\ \ref{sec:espec}) and its interpretation in terms of the GZK-effect suggests that the closest sources of UHECRs are situated within the GZK volume of $d_{\rm GZK} \la 100$\,Mpc. At these scales the matter distribution in the Universe is inhomogeneous, and so must be the distribution of the UHECR sources. If propagation of UHECRs at these distances is quasi-rectilinear, anisotropies would be expected, showing variations at large angular scales and possibly point sources.

\begin{figure}[!t]
\centering
\includegraphics[width=.48\textwidth]{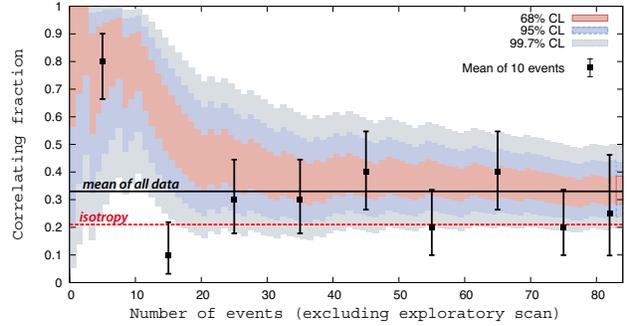}
\vspace*{-2mm}\caption{{\small The most likely value of the degree of correlation $p_{\rm data} = k/N$ is plotted as a function of the total number of time-ordered events (excluding those in period I). The 68\%, 95\% and 99.7\% confidence level intervals around the most likely value are shaded. The horizontal dashed line shows the isotropic value $p_{\rm iso} = 0.21$ and the full line the current estimate of the signal $p_{\rm data} = 0.33\pm0.05$. The black symbols show the correlation fractions bins of independent 10 consecutive events.}}\label{fig:AGN-corr}
\end{figure}

The Pierre Auger Collaboration in fact reported \cite{Abreu-10} directional correlations of UHECR at $E>5.5\cdot 10^{19}$\,eV with AGN from the V\'{e}ron-Cetty-V\'{e}ron catalog \cite{VCV} within 75\,Mpc on an angular scale of $3.1^\circ$ at the 99\,\% CL. The optimal parameters were found using a exploratory scan (Period I) and independent data (Period II) showed 8 of 13 events correlating. An update then yielded 21 of 55 events (Period II+III) correlating for the same parameter set. Here, we present the latest update including data up to June 2011 (c.f.\ Fig.\ \ref{fig:AGN-corr}) which yields a total of 28 of 84 events (Period II+III+IV) showing a correlation on a $3.1^\circ$-scale with a nearby AGN. The overall correlation strength thus decreased from $(69^{+11}_{-13})$\,\% initially to $(33\pm5)$\%. The chance probability of observing such a correlation from a random distribution remains below 1\,\%. Cumulative plots are often misleading and Fig.\,\ref{fig:AGN-corr} may be interpreted as a signal that is fading away. Thus, the superimposed black symbols show in addition the averages of 10 independent consecutive events. Obviously, the first bin is an upwards fluctuation by about $3\sigma$ from the mean of all events while the rest of the dataset does not show any peculiarity. Evidently, more data is needed to arrive at a definite conclusion.

Interestingly, at this meeting the Telescope Array Collaboration presented, though with a much lower exposure of $\sim 15$\% of Auger, an analysis of the northern sky adapting the parameters from the Auger collaboration (for a recent update see \cite{TA-AGN-12}). With 11 correlating events of 25 being above their energy threshold, they find a signal strength of 44\%. Correcting this for the larger chance probability of $p_{\rm iso} =0.24$ compared to 0.21 in Auger, a good agreement of the data sets can be concluded. However, the TA events alone can originate from an isotropic distribution with a chance probability of about 2\,\%. The combined Auger and TA chance probability of observing such a correlation is at the $10^{-3}$ level.

The sky region around Centaurus A is populated by a larger number of high energy events compared to the rest of the sky, with the largest departure from isotropy at $24^\circ$ around the center of Cen A with 19 events observed and 7.6 expected for isotropy. However, a Kolmogorov-Smirnov test shows a chance probability for this to occur at a level of 4\,\%. Similarly, a search for directionally-aligned events (or `multiplets') expected from sets of events coming from the same source after having been deflected by intervening coherent magnetic fields shows one 12-plet with an energy threshold of 20\,EeV. The probability that it appears by chance from an isotropic distribution of events is again 6\,\%. Thus, there is no significant evidence for the existence of correlated multiplets in the present data set \cite[G.\ Golup]{ICRC-Comp-III} and \cite{Abreu-12-Multiplet}. It will be interesting to check if some of the observed multiplets grow significantly or if some new large multiplets appear. If one of them were a real multiplet, doubling the statistics should double its multiplicity, i.e.\ the significance does not increase as $\sqrt{N}$ but much faster.

The Pierre Auger Observatory has sensitivity also to neutron fluxes produced at cosmic ray acceleration sites in the Galaxy. Because of relativistic time dilation, the neutron mean decay length is $(9.2\times E)$~kpc, where $E$ is the neutron energy in EeV. A blind search over the field of view of the Auger Observatory for a point-like excess yields no statistically significant candidates. The galactic center is a particularly interesting target because of the presence of a massive black hole. The results for the window centered on it and for $E \geq 1$~EeV shows no excess with a 95\% CL upper limit on the flux from a point source in this direction of 0.01~km$^{-2}$~yr$^{-1}$ \cite[B.~Rouill\'e d'Orfeuil]{ICRC-Comp-III}, which updates the bounds obtained previously \cite{Auger07}. We note that for directions along the Galactic plane the upper limits are below 0.024~km$^{-2}$~yr$^{-1}$, 0.014~km$^{-2}$~yr$^{-1}$ and 0.026~km$^{-2}$~yr$^{-1}$ for the energy bins $[1-2]$~EeV, $[2-3]$~EeV and $E \geq 1$~EeV, respectively.

A targeted search has also been performed to test potential sources of galactic cosmic rays, such as SNR, pulsars and Pulsar Wind Nebul\ae{} (PWN). The candidate sources are expected to be strong gamma-ray emitters at GeV and TeV energies. 
For this reason, we apply a neutron search also to Galactic gamma-ray sources extracted from the Fermi LAT Point Source Catalog \cite{Abdo10a} and the H.E.S.S. Source Catalog\footnote{\small{http://www.mpi-hd.mpg.de/hfm/HESS/pages/home/sources/}}. Targets were selected among the sources located in the portion of the Galactic plane, defined as $|b|<10^\circ$, covered by the FOV of the SD, and located at a distance shorter than 9~kpc ($\lambda_{\rm n}$ at 1~EeV) \cite[B.~Rouill\'e d'Orfeuil]{ICRC-Comp-III}. The stacked signal computed from the SD data at the positions of the two sets of sources under study and for the same energy bins used in the Galactic plane search has not yet yielded an excess.\\

\begin{figure}[!t]
\centering
\includegraphics[width=.48\textwidth]{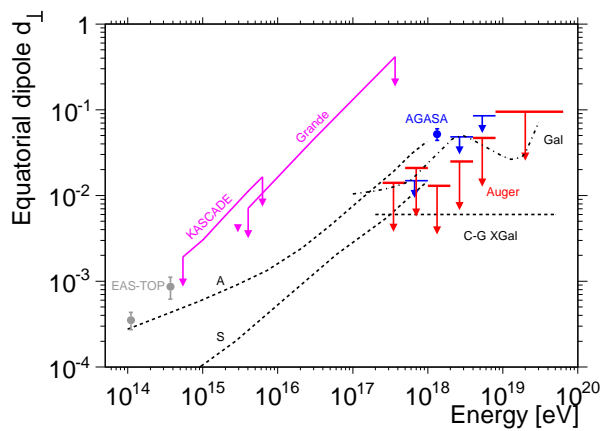}
\vspace*{-2mm}\caption{{\small Upper limits on the anisotropy~: equatorial dipole component $d_\perp$ as a function of energy from Auger. Results from EAS-TOP, AGASA, KASCADE and KASCADE-Grande experiments are also displayed, in addition to several predictions (see \cite[H.~Lyberis]{ICRC-Comp-III} for references.}}
\label{fig:ul}
\end{figure}

Besides searching for point sources of charged cosmic rays or neutrons, the large scale distribution of arrival directions of CRs represents another important tool for understanding their origin. Using data from the SD array, upper limits below 2\% at 99\% $C.L.$ have been recently reported for EeV energies on the dipole component in the equatorial plane~\cite{auger-11}. Such upper limits are sensible, because cosmic rays of galactic origin, while escaping from the galaxy in this energy range, might generate a dipolar large-scale anisotropy with an amplitude at the \% level as seen from the Earth. Even for isotropic extragalactic cosmic rays, a large scale anisotropy may remain due to the motion of our galaxy with respect to the frame of extragalactic isotropy. This anisotropy would be dipolar in a similar way to the Compton-Getting effect~\cite{Compton} in the absence of  the galactic magnetic field.

An update of the results of searches for first harmonic modulations in the right ascension distribution of cosmic rays is presented in Fig.\,\ref{fig:ul} \cite[H.~Lyberis]{ICRC-Comp-III}. The upper limits at  99\% CL obtained here provide the  most stringent bounds at present above $2.5 \times 10^{17}$\,eV. Some predictions for anisotropies arising from models of both galactic and extragalactic cosmic ray origin are included in the plot together with data from other experiments. In models $A$ and $S$ ($A$ and $S$ standing for 2 different galactic magnetic field symmetries~\cite{Candia}), the anisotropy is caused by drift motions due to the regular component of the galactic magnetic field, while in model $Gal$~\cite{calvez}, the anisotropy is caused by purely diffusive motions due to the turbulent component of the field. Some of these amplitudes are challenged by our current sensitivity. For extragalactic cosmic rays considered in model $C$-$G\,Xgal$~\cite{serpico}, the motion of our galaxy with respect to the CMB (supposed to be the frame of extragalactic isotropy) induces the small dipolar anisotropy (neglecting the effect of the galactic magnetic field).

\begin{figure}[!t]
\centering
\includegraphics[width=.4\textwidth]{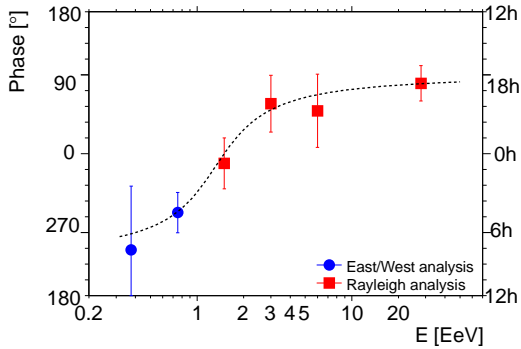}
\vspace*{-3mm}\caption{{\small Phase of the first harmonic as a function of energy. The dashed line, resulting from an empirical fit, is used in the likelihood ratio test (see text) \cite[H.~Lyberis]{ICRC-Comp-III}.}}
\label{fig:phases}
\end{figure}

While the measurements of the amplitudes do not provide any evidence for anisotropy, it is interesting to note that the phase shown in Fig.\,\ref{fig:phases} suggests a smooth transition between a common phase of $\simeq 270^\circ$ below 1~EeV and another phase (right ascension $\simeq 100^\circ$) above 5~EeV. This is potentially interesting, because, with a real underlying anisotropy, a consistency of the phase measurements in ordered energy intervals is indeed expected with lower statistics than that required for the amplitudes to significantly stand out of the background noise. Applying a Likelihood test leads to a probability of $\sim 10^{-3}$ of observing this from a random distribution. However, since we did not perform an \textit{a priori} search for such a smooth transition in the phase measurements, no confidence level can be derived from this  result. 

The infill surface detector array which is now operating at the Pierre Auger Observatory will allow us to extend this search for large scale anisotropies to lower energy thresholds.

\section{Serendipity observations and interdisciplinary science}
\label{sec:serendipity}

The hybrid character of the Auger Observatory, but also the surface and fluorescence detectors themselves allow a number of studies beyond cosmic ray physics. 

A first study, using the so-called scaler mode data of the SD has been presented in \cite{Abreu:2011wh} and is updated at this conference \cite[H.~Asorey]{ICRC-Comp-III}. It uses the count rates of low energy secondary cosmic ray particles (deposited energy $\ga 15$\,MeV and $15 \la E_{\rm dep} \la 100$\,MeV, using two different triggers) recorded continuously for self-calibration purposes for all of the 1660 SD stations. With each detector recording about 3600~Hz, we thus record a total of about 6~MHz so that even very small changes of the rates due to atmospheric and solar changes can be monitored. This enables the SD of Auger to address questions of solar cosmic rays and allows to study Forbush events. A good agreement between neutron monitor and the scaler data is found when accounting for different geomagnetic cut-offs of detectors located at different latitudes and for different effective energy thresholds of neutron monitors and the Auger SD stations~\cite{Abreu:2011wh}.

\begin{figure}[!t]
\centering
\includegraphics[width=.48\textwidth]{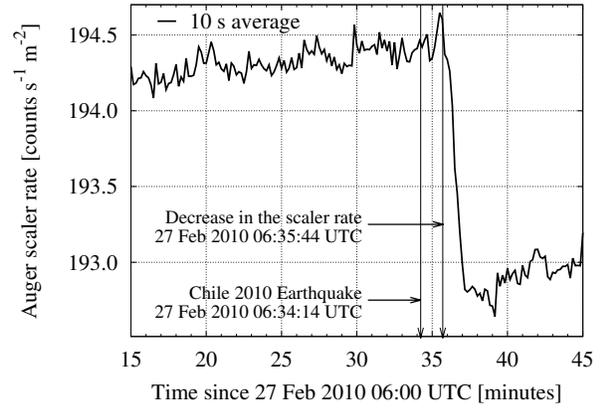}
\vspace*{-2mm}\caption{{\small Ten seconds average of the Auger scaler rate for the 27 Feb 2010 Chile major 8.8 magnitude earthquake. A strong $24\,\sigma$ decrease is found $90\pm 2$\,(stat)\,seconds afterwards, compatible with the time delay expected for seismic S-waves traversing the distance from the epicentre to the Auger
Observatory.}}
\label{fig:quake}
\end{figure}

Instead of using averaged scaler rates for the whole array, it is also possible to study the scaler rate of individual stations, in order to study the propagation of some phenomena across the Auger SD, like the crossing of a storm over the $3000$\,km$^{2}$ of the array. This is because the flux of secondary particles changes as the pressure front moves from across the detector field.  Additional analyses to study the influence of the variation of electric fields on the flux of EAS particles are currently being carried out as well.

Interestingly, the $8.8$ magnitude earthquake in Chile on 27 Feb 2010 06h34 UTC with the epicentre located about 300\,km SW from the Auger Observatory left traces in the SD as well.  The averaged scaler rate for the whole array and also for individual stations showed a $24\,\sigma$ decrease beginning $(90$\,$\pm$\,$2)$\,seconds after the earthquake.  This delay is compatible with the propagation of seismic S-waves over that distance.  The scaler rate from 6h15 to 6h45 UTC is shown in figure \ref{fig:quake}.  Although other minor quakes have been recorded by seismographs near the SD, no other similar effects have been found in 6 years of data. Detailed analyses to identify the causes of the observed drop in the scaler rate are underway.  These include simulations and shaking tests of selected detectors in the array. After 6 hours, the scaler rate recovered to the mean value for February 2010. 

\begin{figure}[!t]
\centering
\includegraphics[width=.48\textwidth]{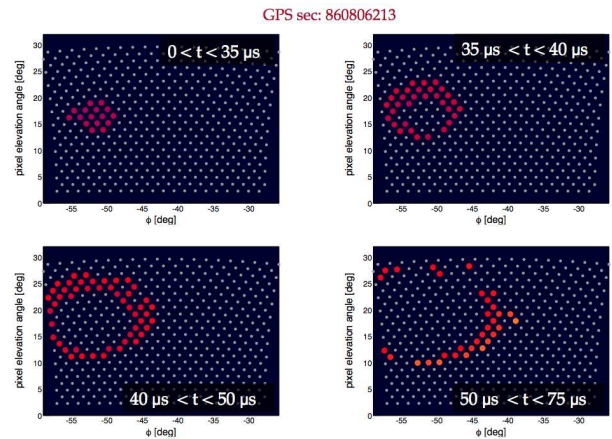}
\vspace*{-2mm}\caption{{\small FD camera image for 4 consecutive time windows as indicated. It shows the time evolution of an elve located at about 80~km altitude at a distance of 580~km from the observatory \cite[A.~Tonachini]{ICRC-Comp-IV}.}}
\label{fig:elve}
\end{figure}

Also quite unexpectedly, during a normal FD data taking shift an unusual event has been observed with a well defined space-time structure: a luminous ring starting from a cluster of pixels, and expanding in all directions \cite[A.~Tonachini]{ICRC-Comp-IV}. Usually, such kind of events lasting for much longer than 70 $\mu$s and with such a high multiplicity are rejected by the T2 trigger because of being caused by lightning with high probability. Due to this rejection, only three of such unusual events were recorded. By careful reconstruction of the timing, these events could be identified as elves originating from lightning in the western part of Argentinia.
Elves are transient luminous phenomena originating in the D-layer of the ionosphere, high above thunderstorm clouds, at an altitude of approximately 90\,km. With a time resolution of 100 ns and a space resolution of about 1 degree, the FD can provide an accurate 3D measurement of elves for thunderstorms which are below the horizon. To  improve the detection efficiency for such kind of interesting and not well understood phenomena, a dedicated trigger will be implemented in the future.

The Auger Observatory allows us to perform a number of further interdisciplinary science studies, mostly related to atmospheric sciences (study of aerosols, atmospheric gravity waves, etc.) but includes also biological studies in the pampa as well as related studies of earthquakes either directly by its instrumentation operated or indirectly by providing infrastructure for non-cosmic ray scientific communities.

\section{Summary and Conclusions}
\label{sec:summary}

The Pierre Auger Observatory has reached a cumulative exposure of more than 25\,000 km$^2$\,sr\,yr by the time of writing this article. This exceeds by far the total statistics recorded by all other observatories. A great deal of new insights are provided by these data, but many new questions have appeared. This is most prominently about the origin of the suppression of the CR flux at highest energies and - related to this - the mass composition and anisotropies at the highest energies. The Observatory will continue to collect data with unprecedented precision for several more years and it is hoped that these data will help to unravel the puzzles about the most energetic particles in nature. 

{\small KHK acknowledges financial support by the German Ministry of Research and Education, by the Helmholtz Alliance for Astroparticle Physics (HAP) and by DAAD.}

\clearpage

\end{document}